\documentstyle[prl,multicol,aps,epsfig]{revtex}
\begin{document}

\title{Reconciling Conductance Fluctuations and \\ the  Scaling
Theory of Localization}

\author{Keith Slevin}
\address{Dept. of Physics, Graduate School of Science,
Osaka University, \\ 1-1 Machikaneyama, Toyonaka,
Osaka 560-0043, Japan}

\author{Peter Marko\v{s}}
\address{Institute of Physics, Slovak Academy of Sciences,
D\'ubravsk\'a cesta 9, 842 28 Bratislava, Slovakia}

\author{Tomi Ohtsuki}
\address{Department of Physics, Sophia University,
Kioi-cho 7-1, Chiyoda-ku, Tokyo 102-8554, Japan}

\maketitle

\begin{abstract}
We reconcile the phenomenon of mesoscopic conductance fluctuations with 
the single parameter scaling theory of the Anderson transition.
We calculate three averages of the conductance distribution: 
$\exp\left(\left<\ln g\right>\right)$, 
$\left< g\right>$ and $1/\left<R\right>$ where $g$ is the
conductance in units of $e^2/h$ and $R=1/g$ is the resistance
and demonstrate that these quantities obey single parameter
scaling laws.
We obtain consistent estimates of the
critical exponent from the scaling of all these quantities.
\end{abstract}


\begin{multicols}{2}

In the original proposal of the scaling theory of localization the
conductance $g$ (in units of $e^2/h$) is the relevant parameter \cite{AALR79}. 
The scaling of this parameter is deduced by looking at the limiting 
cases of a good metal, where weak localization theory is applicable, 
and the strongly localized regime where an exponential dependence
on system size is expected. 
A smooth monotonic interpolation is supposed between the two
limits.
The central equation of the theory is the $\beta$ function
\begin{equation}
\beta(g) \equiv \frac{{\mathrm d} \ln g}{{\mathrm d} \ln L}
\label{betag}
\end{equation}
which describes how the conductance of an $L \times L \times L$ cube
is renormalized with system size $L$.

The scaling theory has been very influential. It is the basis for
the predictions that the Anderson transition is continuous and that
the lower critical dimension for the Anderson transition is two.

Shortly after its proposal, the pioneering numerical calculations 
of Pichard \cite{PS81}, and MacKinnon and Kramer \cite{MKK83} provided 
indirect support for the scaling theory. In these calculations the 
localization length
of electrons on quasi- one dimensional bars was calculated. The theory 
of finite size scaling was then applied to deduce the critical parameters 
from the dependence of the localization length on the transverse dimension 
of the bars. This has proved useful for making quantitative estimates of 
critical parameters \cite{MacK94,SO97,SO99,SOK00,MRSU00}. 

It became clear later, in particular after the investigation 
of universal conductance fluctuations in mesoscopic systems, that the 
conductance of a disordered system is a random variable \cite{A85,LS85}. 
In the vicinity
of the Anderson transition the fluctuations in $g$ are of the same
order of magnitude as it's mean value \cite{SO97}.
The smooth scaling behavior predicted by (\ref{betag}) is clearly
inconsistent with the fluctuating behavior of the conductance which 
occurs in practice and casts strong doubts over the soundness of
the scaling theory \cite{AKL86}. 
There seem to be two principal remedies which we discuss in turn 
\cite{S87,CS92}. 

First, we could attempt to establish that the distribution of 
conductance $p_L(g)$ (in the limit that the system size $L$ and 
the correlation length $\xi$ are much longer than any microscopic 
lengths) obeys a single parameter scaling law. The precise meaning of 
this statement is that it should be possible to parameterize the 
bulk of the conductance distribution with a single parameter.
If we denote this parameter by $X$ then the bulk of the conductance 
distribution must be of the form 
\begin{equation}
p_L(g) \simeq F(g;X(L))
\label{Strong1}
\end{equation}
and the parameter $X$ must obey a single parameter scaling law
\begin{equation}
\frac{{\mathrm d} \ln X(L)}{{\mathrm d} \ln L} = \beta(X) 
\label{Strong2}
\end{equation}
Note that $X$ need not necessarily be one of the moments of the distribution.
Indeed, single parameter scaling of a distribution
does not necessarily imply scaling of its moments. The moments may 
be dominated by non- universal tails of the distribution or might 
not even exist.
We shall refer to this first possibility as strong single parameter 
scaling.

Second, we could attempt to establish that some typical or average
conductance obeys a single parameter scaling law. This is a somewhat 
weaker statement since one parameter scaling of some average or typical
quantity does not imply single parameter scaling of the corresponding
distribution. For example, it might happen that several independent
parameters are needed to describe the distribution.
For this reason we shall refer to this as weak single parameter scaling.

Our purpose in this paper is to firmly establish single parameter scaling
in the weaker sense given above, for
systems near the critical point of the Anderson transition in 
three dimensions.
A secondary objective is to identify which averages or typical values
obey single parameter scaling laws.
By simulating the conductance distribution for large ensembles of
disordered $L \times L \times L$ cubes we have achieved both objectives.

We supposed perfect leads were attached to a pair of opposite sides of
each cube and used the Landauer formula to relate the (Landauer) conductance 
$g_L$ to the transmission matrix $t$ which describes the transmission
of electrons from one lead to the other
\cite{L57}.
\begin{equation}
g_L = 2 {\mathrm tr}~ tt^{\dagger}
\label{gL}
\end{equation}
We use the notation $g_L$ to emphasize that this is the
conductance that would be measured in a two probe measuring
geometry. In the original work on the scaling theory the Thouless
conductance $g=E_C/\Delta$ was considered where $E_c$ is the
Thouless energy and $\Delta$ the mean energy level spacing. The 
Landauer conductance and the Thouless conductance are not 
completely equivalent. Their relation has been considered in detail 
by Braun {\it et al} \cite{BHMKM97} who state that the contact 
resistance, which 
is always present in a two terminal measurement, should be subtracted 
from $g_L$ . Hence we study the statistics of
\begin{equation}
\frac{1}{g} = \frac{1}{g_L} - \frac{1}{2N}
\end{equation}
Here $N\equiv N(E_F)$ is the number of propagating channels
in the contacts at Fermi energy $E_F$ and $1/2N$ is the contact
resistance appropriate for the situation we have simulated.

The motion of the electrons in the system is described by the 
Anderson model
\begin{equation}
 H = V \sum_{<i,j>} C_i^{\dagger}C_j +
     \sum_i W_i C_i^{\dagger}C_i ,
\label{Hamiltonian}
\end{equation}
where $C_i^{\dagger}(C_i)$ denotes the creation (annihilation)
operator of an electron at the site $i$ of a 3D cubic lattice.
The amplitude of the random potential at site $i$ is $W_i$.
Hopping is restricted to nearest neighbors and its
amplitude is taken as the unit of energy, $V=1$.
We supposed a box distribution with each $W_i$ uniformly distributed 
on the interval $[-W/2,W/2]$. Previous work has verified the universality
of the critical behavior in this model with respect to the choice
of distribution of the random potential \cite{SO99,SO01,M99}.  
We imposed fixed boundary conditions in the transverse directions
since we have found in previous work that corrections to scaling
vanish more quickly with system size in this case \cite{SOK00,SWS99}.
The Hamiltonian (\ref{Hamiltonian}) has both time reversal and spin 
rotation symmetries
so that the observed critical behavior should be that of the 
orthogonal universality class.
We used the method of Pendry {\it et al} \cite{PMKR92} 
to calculate the
transmission matrix which appears in (\ref{gL}). We set the Fermi 
energy $E_F=0.5V$ and for each combination of disorder $W$ and system
size $L$ we generated an ensemble of $1,000,000$ samples (except for 
$L=16$ where 500,000 were generated). 
This allowed us to estimate the various averages of the conductance
to roughly an accuracy of $0.1\%$.

We examined the behavior of three different averages
of the conductance distribution: $\exp\left( \left<\ln g\right> \right)$, 
$\left< g\right>$ and $1/\left< R\right>$ 
where $R=1/g$ is the resistance. Note that for each average it is possible 
to define a different $\beta$ function. For example, for 
the typical conductance $\exp\left( \left<\ln g\right> \right)$
\begin{equation}
\beta(\exp\left(\left<\ln g\right>\right)) \equiv \frac{{\mathrm d} \left< \ln g \right> }{{\mathrm d} \ln L}.
\end{equation}
In the critical region the conductance fluctuations are of the same order
of magnitude as the mean conductance so that these averages are not at
all equivalent.

We fitted the disorder and system size dependence of the averages to 
the standard scaling
form. Taking the typical conductance as an example we supposed that
\begin{equation}
\left< \ln g \right> = F \left( \psi L^{1/\nu} , \phi L^y \right),
\end{equation}
where $\psi$ is a relevant scaling variable and $\phi$ is an 
irrelevant scaling variable which allows us to take account of
corrections to scaling. Such deviations from perfect scaling
are always present in a simulation of a finite system and it is
necessary to have some means of accounting for them.
We approximated this scaling function by its first order expansion
in the irrelevant scaling variable and fitted the numerical data
to the form
\begin{equation}
\left< \ln g \right> = F_0(\psi L^{1/\nu}) + \phi L^y F_1 (\psi L^{1/\nu}).
\end{equation}
The scaling variables were approximated by expansions
in terms of the dimensionless disorder $w=(W_c-W)/W_c$ where
$W_c$ is the critical
disorder separating the insulating and metallic phases.
\begin{equation}
\psi = \psi_1 w + \psi_2 w^2, \ \ \ \ \ \phi=\phi_0.
\end{equation}
The critical exponent $\nu$ describes the divergence of the 
localization (correlation) near the critical point
\begin{equation}
\xi = \xi_0 \left| \psi \right|^{-\nu}.
\end{equation}
The absolute scale of the localization length $\xi_0$ 
cannot be determined from this fit.
The system size dependence of the irrelevant scaling variable 
is described by an exponent $y<0$.
The functions
$F_0$ and $F_1$ were expanded to third order in $w$.
We found that this fitting scheme was the simplest
which still
allowed for goodness of fit probabilities in excess of $0.1$.

\begin{figure}
\epsfig{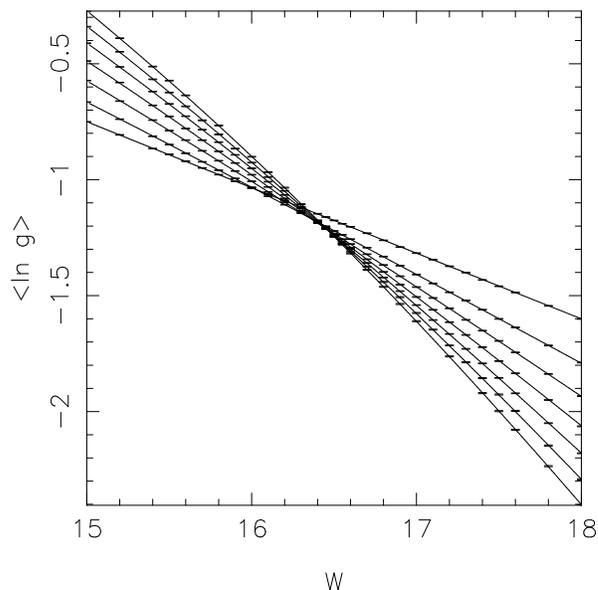}
\caption{The logarithm of the typical conductance versus the
amplitude of the potential fluctuations for system sizes $L=4,6,8,
10,12,14$ and $16$. The solid lines are the fit to the data.}
\label{F1}
\end{figure}

\begin{figure}
\epsfig{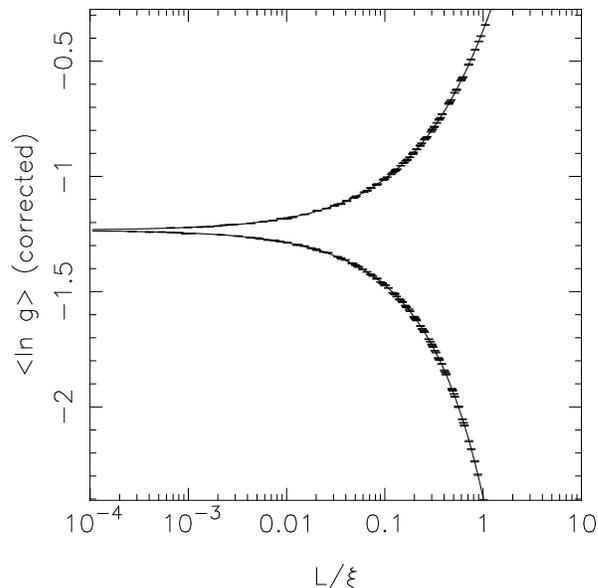}
\caption{The same data as in Fig. 1 after corrections to scaling are
subtracted and plotted versus $L/\xi$ to exhibit the single
parameter scaling function.}
\label{F2}
\end{figure}

\begin{figure}
\epsfig{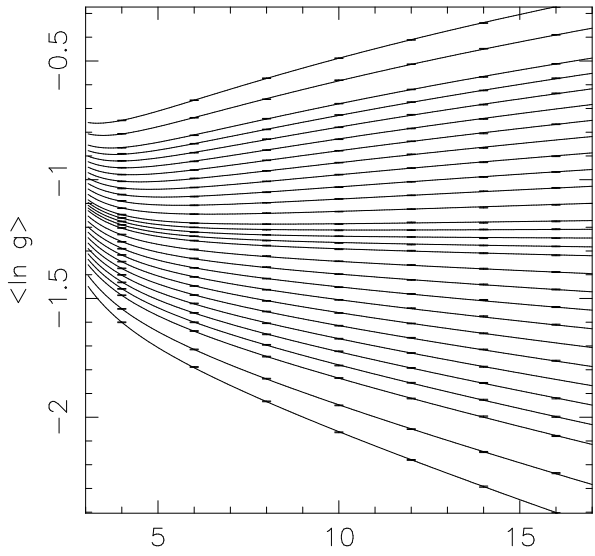}
\caption{The same data as in Figure 1 but plotted versus system size.}
\label{F3}
\end{figure}

\begin{figure}
\epsfig{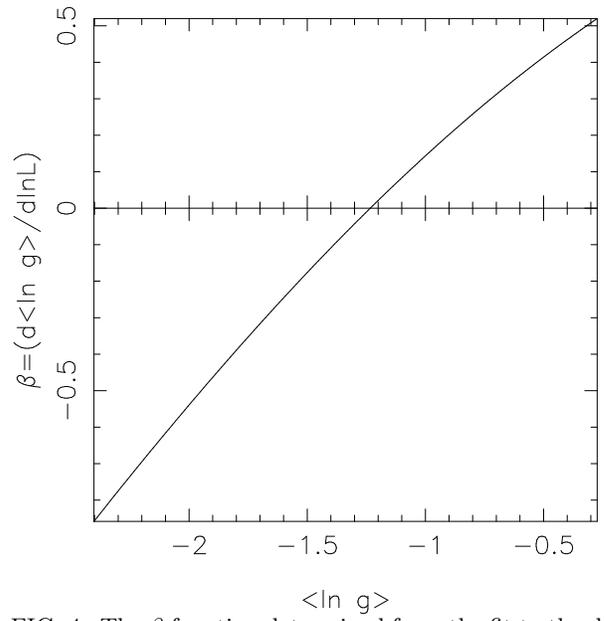}
\caption{The $\beta$ function determined from the fit to the data in Fig. 1.}
\label{F4}
\end{figure}

The results of the analysis are displayed in Table \ref{TABLE1} and 
Figures \ref{F1}-\ref{F4}.
A number of points can be noticed.
First, acceptable fits to the single parameter scaling law are obtained
for all three averages.
Second, mutually consistent values for the critical exponent are 
obtained and these estimates are also consistent with estimates based
on the scaling of the localization length of electrons on bars
\cite{SO99}.
Third, mutually consistent values of the critical disorder are
obtained.
Finally, the critical values of the quantities considered vary widely,
indicative of the broadness of the distribution of the conductance
fluctuations in the critical regime. This means that the precise form of the
$\beta$ function is different for each average, though all should have 
the same slope $(=1/\nu)$ at their respective zeros \cite{S87}.
Note that the $\beta$ function will also depend on the boundary
conditions since it is known that critical conductance distribution 
depends strongly on the boundary conditions even in the limit that
$L \rightarrow \infty$ \cite{SOK00}.

These results firmly establish that the typical conductance, the mean 
conductance and the mean resistance all obey single parameter scaling
in the critical regime.

To what extent do these results also support scaling 
in the strong sense? Let us consider the vicinity of the
metallic, insulating and critical fixed points in turn. 

The metallic fixed point can be reached by taking the limit 
$L\rightarrow\infty$ with $\xi$ fixed from any starting point on the
metallic side of the transition.
In this limit $p_L(g)$ approaches a normal distribution with a
size independent variance \cite{S87}. 
Therefore a single parameter, $\left< g \right>$, is sufficient to 
parameterize the distribution. Further it can be established that
$\left< g\right>$ obeys a single parameter scaling law using weak localization
theory near the metallic fixed point \cite{AALR79}.

The insulating fixed point can be reached by taking the limit 
$L\rightarrow\infty$ with $\xi$ fixed from any starting point 
on the insulating side of the transition. In the insulating regime 
exact results are available only for one dimensional systems \cite{P95}. 
It is found that distribution of conductance is log- normal with a 
width which is related to the mean of the distribution (provided
certain conditions are satisfied \cite{DLA00}.) Hence a single parameter 
$\left< \ln g\right>$ is sufficient to parameterize the distribution.
Also $\left< \ln g\right>$ obeys a single parameter
scaling law in the limit $L\gg\xi$.
Calculations of the average conductance and the average resistance, 
in this limit, in strictly one dimension, give
\begin{equation}
\left<g \right> = \frac{1}{2} \left( \frac{\pi \xi}{2 L} \right)^{3/2}
\exp\left(-L/2 \xi \right)
\end{equation}
and
\begin{equation}
\left<R \right> = \frac{1}{2} \exp\left(4 L/\xi \right)
\end{equation}
showing that both these quantities also obey single 
parameter scaling laws. 
However, the average conductance (and also its higher moments)
are not consistent with the 
log- normal form of the bulk of the conductance distribution. 
They turn out to be determined by the probability of the
occurrence of rare resonance
states with anomalously large conductances which are not described by the
the log- normal distribution\cite{P95}. It is possible, but not certain, that
these results are qualitatively correct for the insulating fixed point 
in three dimensions.

At the critical fixed point $\xi$ diverges and the single parameter
scaling hypothesis predicts a scale independent universal critical 
conductance distribution $p_c(g)$.
This has been confirmed in numerical simulations 
\cite{SO97,SOK00,SWS99,MK93,M94,M99}. 

Thus, it seems likely that the conductance distribution also obeys 
single parameter scaling in the strong, as well as the weak, senses.
However, the available results are not completely conclusive. It seems
to us that a demonstration that $p_L(g)$ obeys 
$(\ref{Strong1})$ and $(\ref{Strong2})$ in the limit 
$L\rightarrow \infty$ and $\xi \rightarrow \infty$ with $L/\xi$ 
fixed for any value of the ratio $L/\xi$ is required. 

We would like to thank the ISSP and the Slovak Academy of Sciences 
for the use of their computer facilities. 
PM would like to thank the Japan Society for the Promotion of
Science and Sophia University for their hospitality and VEGA for
financial support.

\begin{table}
\begin{tabular}{|l|l|l|l|} \\ 
$X$	   & $\exp\left( \left<\ln g\right> \right)$ & 
$\left< g\right>$  & $1/\left< R\right>$ \\ \hline
$\nu$  & 1.57(56,58)    & 1.58(57,60)     &  1.54(53,56)    \\ \hline
$W_c$  & 16.48(47,49)   & 16.47(45,48)    &  16.49(48,50)   \\ \hline
$X_c$  & 0.291(290,293) & 0.573(570,576)  &  0.100(099,101) \\ \hline
$Q$    & 0.5           & 0.4             &  0.3            \\  
\end{tabular} 
\caption{Results of the scaling analysis for each average $X$.
Estimates of the critical exponent $\nu$ and the critical
disorder $W_c$ together with $95\%$ confidence intervals are given.
$X_C$ is the value of the relevant statistic at the critical point.
The number of data values was $199$ and the number of parameters
in each fit was $12$.
The goodness of fit probability $Q$ is also given.}
\label{TABLE1}
\end{table}

\end{multicols}

\end{document}